\documentclass[prd,twocolumn,superscriptaddress,amsfonts,amssymb,amsmath,showpacs]{revtex4-2}
\usepackage{bm}
\usepackage{amsfonts}
\usepackage{latexsym}
\usepackage[latin1]{inputenc}
\usepackage{graphicx}
\usepackage{amsmath}
\usepackage{relsize}
\usepackage{palatino}
\usepackage{mathpazo}
\usepackage{textcomp}
\usepackage{float}
\usepackage{booktabs}
\usepackage{dcolumn}
\usepackage{booktabs}
\usepackage{multirow}
\usepackage{hyperref}
\hypersetup{
    colorlinks=true,
    linkcolor=purple,
    filecolor=magenta,      
    citecolor=blue
}
\usepackage{amsmath}
\usepackage{xcolor}
\usepackage{orcidlink}
\usepackage[caption=false]{subfig}
\usepackage{commath}
\usepackage[many]{tcolorbox}  
\newtcolorbox{boxB}{
    fontupper = \bf, % font color
    boxrule = 1.5pt,
    colframe = black,
    colback = white,
    rounded corners,
    arc = 5pt   % corners roundness
}
\makeatletter
\long\def\dddddot#1{%
  {\mathop {#1}\limits ^{\vbox to-1.4\ex@ {\kern -\tw@ \ex@ \hbox {\normalfont .....}\vss }}}%
}
\long\def\multidots#1#2{%
  \count@=0
  {{\mathop {#2}\limits ^{\vbox to-1.4\ex@ {\kern -\tw@ \ex@ \hbox {\normalfont %
  \loop%
  \ifnum#1>\count@%
  .%
  \advance\count@ by1%
  \repeat%
  }\vss }}}}%
}
\makeatother

\begin{document}

\title{\bf Constrained cosmological model in $f(Q,T)$ gravity with non-linear non-metricity}

\author{Rahul Bhagat\orcidlink{0009-0001-9783-9317}}
\email{rahulbhagat0994@gmail.com}
\affiliation{Department of Mathematics, Birla Institute of Technology and
Science-Pilani,\\ Hyderabad Campus, Hyderabad-500078, India.}
\author{S.A. Narawade\orcidlink{0000-0002-8739-7412}}
\email{shubhamn2616@gmail.com}
\affiliation{Department of Mathematics, Birla Institute of Technology and
Science-Pilani,\\ Hyderabad Campus, Hyderabad-500078, India.}
\author{B. Mishra\orcidlink{0000-0001-5527-3565}}
\email{bivu@hyderabad.bits-pilani.ac.in}
\affiliation{Department of Mathematics, Birla Institute of Technology and
Science-Pilani,\\ Hyderabad Campus, Hyderabad-500078, India.}
\author{S.K. Tripathy\orcidlink{0000-0001-5154-2297}}
\email{tripathy\_sunil@rediffmail.com}
\affiliation{Department of Physics, Indira Gandhi Institute of Technology, Sarang, Dhenkanal, Odisha-759146, India.}

\begin{abstract}
The $f(Q,T)$ cosmological model has emerged as a promising framework for understanding various aspects of cosmic evolution. In this study, we focused on obtaining the constraints of the free parameters in the non-linear form of non-metricity in $f(Q,T)$ gravity using the $Hubble$, $Pantheon$, and $BAO$ datasets. To determine the best-fit values for the model parameters and the equation of state (EoS) parameter, we employed an MCMC analysis. By examining the error bar plots, we observed that both the model curve and the $\Lambda$CDM curve successfully passed through the range obtained from the datasets. Additionally, we studied the state finder diagnostics and energy conditions to gain insights into the properties of the model. Furthermore, we conducted an analysis using the $Om(z)$ diagnostic, which provides a null test for the validity of the $\Lambda$CDM model.
\end{abstract}

\maketitle
%\textbf{PACS number}: 04.50kd.\\

\section{Introduction} 
The accelerated expansion of the Universe has emerged as a prominent issue in modern cosmology \cite{Riess1998,Perlmutter1999} in recent times. There have been extensive research to investigate whether modifications to general relativity (GR) can account for this accelerated expansion \cite{Capozziello2011,Aghanim2020}. To tackle this issue various hypotheses in different perspectives, have been proposed. One approach is with the metric formalism that involves fixing the Levi-Civita connection and modifying the action with respect to the metric. Another approach is to consider the metric-affine formalism \cite{Hehl1995}. In the framework of Riemannian geometry, generalizing the Einstein-Hilbert action by replacing the Ricci scalar $R$ with an arbitrary function $f(R)$ give rise to $f(R)$ gravity \cite{Nojiri2011,Sotiriou2010}. Similar to the Levi-Civita connection that represents curvature in GR, one can also modify in the gravitational sector to get the teleparallel equivalent of GR (TEGR), where the Weitzenb$\ddot{o}$ck connection represents torsion in teleparallelism \cite{Weitzenboock1923}. In TEGR framework, the $f(\mathcal{T})$ gravity is an explicit modification where the matter Lagrangian becomes a free function of the torsion scalar $\mathcal{T}$ \cite{Ferraro2007}. Further, Weyl proposed an extension of Riemannian geometry where the electromagnetic field arises from the non-metricity of space time, leading to the first unified theory of gravity and electromagnetism \cite{Weyl1918}. The symmetric teleparallel representation provides a third generalization of GR. In this formulation, the non-metricity, expressed by the geometric variable $Q$, governs the properties of the gravitational interaction. The $f(Q)$ gravity is an extension of the non-metricity gravity \cite{Jimenez2018}. The $f(Q)$ gravity theory has shown promising results in the cosmological phenomenology both at the background and perturbation levels \cite{Lu2019,Lazkoz2019,Jimenez2020,Ayuso2021,Hu2022,Dimakis2022,Esposito2022,Khyllep2021,Khyllep2023,Narawade2022a}. Also, it has been successful in confrontation with various observational datasets such as, CMBR, SNIa, BAO, redshift space distortion, etc \cite{Soudi2019,Barros2020,Anagnostopoulos2021,Atayde2021,Frusciante2021,Narawade2022b}.
 
 Further replacing the Lagrangian with an arbitrary function $f$ of the non-metricity $Q$ and the trace of the energy-momentum tensor $T$, the $f(Q,T)$ gravity can be formulated \cite{Xu2019}. One of the notable characteristics of $f(Q,T)$ gravity is the violation of the conservation of energy principle in its standard form. This is because of the non-minimal coupling between matter and curvature in the Lagrangian, which leads to energy exchange between these components \cite{Harko2011}. However, modifications to the theory have been proposed to address this issue, such as the introduction of a coupling term that ensures the conservation of energy in the field equations \cite{Xu2019}. The $f(Q,T)$ gravity presents a promising framework for studying the dynamics of the Universe at cosmological scales. References have shown that this theory offers a new perspective on inflation in the early Universe, addressing the observed power spectrum of cosmic microwave background radiation and overcoming certain limitations of the standard inflationary paradigm \cite{Shiravand2022, Bourakadi2023}. Moreover, it has been demonstrated that $f(Q,T)$ gravity can account for the late-time acceleration of the Universe without the need for dark energy \cite{Koussour2022a,Koussour2022b,Narawade2023a}. In $f(Q,T)$ gravity, the evolution of density parameters has been explored through dynamical system analysis \cite{Pati2023} whereas the possibility of occurrence of future singularities late time behaviour are shown in Refs. \cite{Pati2021a,Pati2021b}. The cosmological perturbation theory that reveals the significant impact of non-minimal couplings between matter and curvature perturbations on the evolution of the Universe has been shown in \cite{Najera2022}. Transit model aligning with the observed value of deceleration parameter has been shown in \cite{Zia2021} whereas the comparison with $\Lambda$CDM model has been performed in \cite{Godani2021}. The accelerating $f(Q,T)$ gravity model has been constrained Using the dynamical system analysis in \cite{Narawade2023}.

 The paper is organised as follows: In Sec. \ref{Sec:II} we have discussed the field equations of $f(Q,T)$ gravity field equations in the spatially flat FLRW space time. The parameterization of Hubble parameter and the cosmological data analysis have been performed in Sec. \ref{Sec:III} using the $Hubble$, $Pantheon$, and $BAO$ datasets with the Markov chain Monte Carlo (MCMC) procedure. In Sec. \ref{Sec:IV}, the cosmographic parameters and the energy conditions have been analysed with the obtained best fit values of the model parameters. The results obtained and the conclusions are given in Sec. \ref{Sec:V}.

\section{Field equations of $f(Q, T)$ theory}\label{Sec:II}
The action of $f(Q,T)$ gravity is given as \cite{Xu2019},
\begin{equation} 
S=\int\left[\dfrac{1}{16\pi}f(Q,T)\sqrt{-g}~d^{4}x+\mathcal{L}_{m}\sqrt{-g}~d^{4}x\right],\label{eq.1}
\end{equation}
where $\mathcal{L}_{m}$ be the matter Lagrangian and $g=det(g_{\mu\nu})$ be the determinant of the metric tensor $g_{ij}$. The non-metricity, $Q\equiv -g^{\mu \nu}( L^k_{~l\mu}L^l_{~\nu k}-L^k_{~lk}L^l_{~\mu \nu})$, where $L^k_{~l\gamma}\equiv-\frac{1}{2}g^{k\lambda}(\bigtriangledown_{\gamma}g_{l\lambda}+\bigtriangledown_{l}g_{\lambda \gamma}-\bigtriangledown_{\lambda}g_{l\gamma})$. The field equations of $f(Q,T)$ gravity can be obtained by varying the action as,
\begin{multline}\label{eq.2}
-\frac{2}{\sqrt{-g}}\bigtriangledown_{k}(F\sqrt{-g}P^{k}_ {~\mu\nu})-F(P_{\mu kl} Q^{\;\;\; kl}_{\nu}-2Q^{kl}_{\;\;\;\mu} P_{kl\nu})\\
-\frac{1}{2}fg_{\mu \nu} = 8 \pi T_{\mu \nu}\left(1-\kappa\right)-8\pi\kappa \Theta_{\mu \nu}.
\end{multline}
We denote $f\equiv f(Q,T)$ and $F=\frac{\partial f}{\partial Q}$, $8\pi\kappa = f_{T}=\frac{\partial f}{\partial T}$. We can obtain the trace of the non-metricity, the super potential of the model and the energy momentum tensor as,

\begin{eqnarray}\label{eq.3}
Q_{k} &=& Q_{k}^{\;\;\mu}\;_{\mu},~~~~~~~ \tilde{Q}_{k}=Q^{\mu}\;_{k\mu},\nonumber\\
P^{k}_{~\mu\nu} &=& -\frac{1}{2}L^{k}_{~\mu \nu}+\frac{1}{4}\left(Q^{k}-\tilde{Q}^{k}\right)g_{\mu \nu}-\frac{1}{4}\delta^{k}_{(\mu}Q_{\nu)},\nonumber\\
T_{\mu \nu}&=&\frac{-2}{\sqrt{-g}} \frac{\delta(\sqrt{-g}L_{m})}{\delta g^{\mu \nu}},~~~~~~~\Theta_{\mu \nu}=g^{kl}\frac{\delta T_{kl}}{\delta g^{\mu \nu}}, 
\end{eqnarray}

The matter-energy-momentum tensor divergence can be described as,
\begin{eqnarray}\label{eq.4}
&&\mathcal{D}_{\mu }T_{\ \ \nu }^{\mu }=\frac{1}{f_{T}-8\pi }
\Bigg[-\mathcal{D}_{\mu }\left( f_{T}\Theta _{\ \ \nu}^{\mu}\right) -\frac{16\pi}{\sqrt{-g}}\nabla _{k}\nabla _{\mu}H_{\nu}^{\ \ k\mu } 
\notag \\
\hspace{-0.5cm} &&+8\pi \nabla _{\mu }\bigg(\frac{1}{\sqrt{-g}}\nabla
_{k}H_{\nu }^{\ \ k\mu }\bigg)-2\nabla _{\mu }A_{\ \ \nu }^{\mu}+\frac{1}{2}f_{T}\partial _{\nu }T\Bigg],
\end{eqnarray}

where $\mathcal{D}_{\mu }$ and $\Theta _{\ \ \nu }^{\mu }$ respectively represents the covariant derivative and canonical energy-momentum tensor. The term $H_{l}^{\ \ \mu \nu}$ is the hyper-momentum tensor density and defined in terms of $f_{T}$ and $\delta T/\delta \widetilde{\Gamma}_{\ \ \mu \nu }^{l}$ and $\delta \sqrt{-g}\mathcal{L}_{M}/\delta \widetilde{\Gamma }_{\ \ \mu \nu}^{l}$. To note $\delta T/\delta \widetilde{\Gamma}_{\ \ \mu \nu }^{l}$ is the derivative of $T$ with respect to the affine connection, and $\delta \sqrt{-g}\mathcal{L}_{M}/\delta \widetilde{\Gamma }_{\ \ \mu \nu}^{l}$ is the derivative of the matter Lagrangian with respect to the affine connection. So,

\begin{equation}\label{eq.5}
H_{l}^{\ \ \mu \nu}\equiv \frac{\sqrt{-g}}{16\pi }f_{T}\frac{\delta T}{\delta \widetilde{\Gamma}_{\ \ \mu \nu}^{l}}+\frac{\delta \sqrt{-g}\mathcal{L}_{M}}{\delta \widetilde{\Gamma}_{\ \ \mu \nu }^{l}}.
\end{equation}

The divergence of the energy-momentum tensor gives rise to the energy balance equation and the momentum conservation equation respectively as,

\begin{equation}
\dot{\rho}+3H(\rho+p) = B_{\mu}u^{\mu}, \label{eq.6}
\end{equation}
and
\begin{equation}
\frac{d^{2}x^{\mu}}{ds^{2}}+\Gamma^{\mu}_{kl}u^{k}u^{l} = \frac{h^{\mu\nu}}{\rho+p}(B_{\nu}-\mathcal{D}_{\nu}p),\label{eq.7}
\end{equation}
An over dot denotes the derivative with respect to time. The term $B_{\mu}u^{\mu}$ in Eq. \eqref{eq.6} accounts for energy creation or annihilation. If $B_{\mu}u^{\mu}=0$ holds at all points in space time, the total energy of the gravitational system is conserved. However, when $B_{\mu}u^{\mu}\neq0$, energy transfer processes or particle production occurs in the system. By varying the gravitational action with respect to the connection, the field equations of $f(Q,T)$ theory can be obtained as,

\begin{eqnarray}\label{eq.8}
\nabla_{\mu}\nabla_{\nu}\bigg( \sqrt{-g}FP^{\mu\nu}_{\ \ \ \ k}+4\pi H_{k}^{\ \ \mu \nu} \bigg)=0,
\end{eqnarray}
These field equations govern the behavior of the gravitational field in the $f(Q,T)$ theory. We consider an isotropic and homogeneous FLRW space time as, 
\begin{eqnarray}\label{eq.9}
ds^{2}=-N^{2}(t)dt^{2}+a^{2}(t)(dx^{2}+dy^{2}+dz^{2}),
\end{eqnarray}
where  $N(t)$ and $a(t)$ respectively denotes the lapse function and scale factor; $\tilde{T}=\frac{\dot{N}(t)}{N(t)}$ is the dilation rate. The non-metricity, $Q=6\frac{H^{2}}{N^{2}}$, where $H=\frac{\dot{a}}{a}$ be the Hubble parameter. For the flat FLRW space time, $Q=6H^2$ and $\tilde{T}=0$. We consider the energy-momentum tensor corresponding to a perfect fluid distribution as, $T^{\mu}_{\nu}=diag(-\rho,p,p,p)$, where $\rho$ denotes the energy density and $p$ represents the pressure. The canonical energy-momentum tensor is $\Theta^{\mu}_{\nu}=diag(2\rho+p,-p,-p,-p)$. Now the field equations of $f(Q,T)$ gravity given in Eq. \eqref{eq.2} in FLRW space time can be obtained as \cite{Xu2019, Pati2021a},
\begin{eqnarray}
p&=&-\frac{1}{16\pi}\left[f-12FH^2-4\dot{\chi}\right],\label{eq.10} \\
\rho&=&\frac{1}{16\pi}\left[f-12F H^2-4\dot{\chi}\kappa_1\right],\label{eq.11}
\end{eqnarray}
where $\kappa_1=\frac{\kappa}{1+\kappa}$ and $\dot{\chi}=F\dot{H}+\dot{F}H$.
 
\section{Hubble Parametrization}\label{Sec:III}
We wish to investigate the $f(Q,T)$ gravity cosmological model using various observational datasets. We shall employ the MCMC analysis with the available \textit{emcee} package (Ref. \cite{Foreman-Mackey2013}), to constrain the parameters of the model. This involves the  exploration of the parameter space through conservative priors and investigating the posterior distributions to obtain one-dimensional and two-dimensional distributions. The one-dimensional distributions represented the posterior distributions of individual parameter, while the two-dimensional distributions showed the covariance between different parameter pairs. We shall  determine $1\sigma$ and $2\sigma$ confidence levels to quantify uncertainty and compare different datasets. For this we need to have the Hubble parameter $H(z)$, which may be derived by assuming some form of the function $f(Q,T)$. Here we consider a form of $f(Q,T)$ with the non-linear non-metricity as,  
\begin{equation}\label{eq.12}
f(Q,T)= -Q+\lambda Q^{n+1}+2~\zeta~T. 
\end{equation}
Subsequently,
\begin{equation*}
F=-1+\lambda (n+1) Q^{n},~~~f_{T}=2\zeta,~~~\kappa=\frac{\zeta}{4\pi},~~~\kappa_{1}=\frac{\zeta}{4\pi + \zeta}    
\end{equation*}
and
\begin{equation}\label{eq.13}
\dot{\chi}=\left[-1+\lambda (n+1)(2n+1)Q^{n}\right]\dot{H}.
\end{equation}
From Eq. \eqref{eq.6}, one can retrieve the evolution equation of the Hubble function as,
\begin{equation}\label{eq.14}
\dot{\chi}-4\pi(p+\rho)(1+\kappa)=0
\end{equation}
Substituting $p=\omega \rho$ and $\rho=3H^2\Omega$ in Eq. \eqref{eq.14}, we get
\begin{equation}\label{eq.15}
\big[-1+\lambda (n+1)(2n+1)Q^n\big]\dot{H}=12\pi H^{2}\Omega(1+\omega)(1+\kappa),
\end{equation}
where $\omega$ and $\Omega$ respectively represents the equation of state (EoS) and denity parameter. Now, the Hubble parameter and redshift can be expressed as, $\dot H=\frac{dH}{dt}=-(1+z)H(z)\frac{dH}{dz}$. Hence Eq. \eqref{eq.15} becomes,
\begin{multline}\label{eq.16}
\left[\frac{1}{H}-\lambda (n+1)(2n+1)6^{n}\frac{H^{2n}}{H}\right]dH\\
=12 \pi \Omega(1+\omega)(1+\kappa)\frac{dz}{1+z}~,
\end{multline}
Taking integration on both sides and considering the linear approximation for $e^x \approx 1+x$, we can obtain the Hubble parameter as,
\begin{equation}\label{eq.17}
H(z)=H_{0}\left[\frac{1+(1+z)^\alpha}{1-\beta}\right]^{\frac{1}{2n}},
\end{equation}
where
\begin{eqnarray*}
\alpha &=& 24 \pi n \Omega(1+\omega)(1+\kappa), \nonumber \\    
\beta &=& \lambda (n+1)(2n+1)6^{n}
\end{eqnarray*}

We shall now introduce the cosmological dataset that will be utilized in the MCMC analysis such as $Hubble$ data, $Pantheon$ observations complemented by $BAO$ data.\\

\noindent{\bf Cosmic Chronometers:}  The cosmic chronometer method involves spectroscopic dating techniques applied to passively-evolving galaxies which allows to directly measure the Hubble function at various redshifts up to $z\approx 2$. These measurements are independent of specific cosmological models. However, they still rely on accurate modeling of stellar ages using robust stellar population synthesis techniques. The method involves comparing the age difference between two passively evolving galaxies at different redshifts. This enables the inference of $\Delta z/\Delta t$ from observations, which in turn allows for the computation of $H(z) = -(1+z)^{-1}\Delta z/\Delta t$. Consequently, cosmic chronometers (CC) have been considered to be more reliable than other methods based on determining the absolute age of the galaxies \cite{Jimenez2001}.
 \begin{equation}\label{eq.18}
 \chi^{2}_{CC}=\mathlarger{\mathlarger{\sum}}_{i=1}^{32}\frac{\big[H(z_{i},\Theta)-H_{obs}(z_i)\big]^2}{\sigma_H^2(z_i)},
 \end{equation}
 where $H(z_i,\Theta)$ represents the Hubble parameter with the model parameters, $H_{obs}(z_i)$ represents the observed Hubble parameter values and $\sigma_H^2(z_i)$ is the standard error. we shall use a set of thirty-two cosmic chronometer data points \cite{Moresco2022}.\\

\noindent{\bf $Pantheon$ Dataset:} The Pantheon compilation, a comprehensive dataset for studying Type Ia supernovae, integrates data from various surveys and telescopes. This compilation encompasses $1048$ data points spanning a redshift range of $0.01$ to $2.3$ \cite{Scolnic2018}. It incorporates observations from renowned projects such as the Super-Nova Legacy Survey (SNLS), Sloan Digital Sky Survey (SDSS), Hubble Space Telescope (HST) survey, and Panoramic Survey Telescope and Rapid Response System (Pan-STARRS1). Combining the information from all these diverse sources, the Pantheon compilation offers valuable insights into the properties and behavior of Type Ia supernovae and their cosmological significance. The focus is to estimate the model parameters by comparing the observed and theoretical values of the distance moduli, which can be defined as,
\begin{equation}\label{eq.19}
\mu(z_{i},\Theta)=m-M=5\log_{10}\big[D_{L}(z_{i},\Theta)\big]+\mu_0,
\end{equation}
where $\mu_{0}$ be the nuisance parameter. The dimensionless luminosity distance $d_{L}$ given as,
\begin{equation}\label{eq.20}
D_{L}(z_{i},\Theta) = (1+z_{i})\int_{0}^{z_{i}}\frac{d\tilde{z}}{H(\tilde{z},\Theta)},
\end{equation}
Further, using an arbitrary fiducial absolute magnitude $M$, the apparent magnitude of each Type Ia supernova to be calibrated. This approach involves the use of theoretical models to predict the distance modulus for a given set of cosmological parameters. Then comparing these predictions to the observed values for the supernovae in the $Pantheon$ catalog. After that the cosmological parameters are constrained by minimizing the $\chi^2$ likelihood \cite{Conley2011} as,
\begin{equation*}
\chi^2_{SN} = \big(\Delta\mu(z_i, \Theta)\big)^T C^{-1}\big(\Delta\mu(z_i, \Theta)\big),
\end{equation*}
where $\Delta\mu(z_i, \Theta) = \left(\mu(z_i, \Theta) - \mu_{obs}(z_i)\right)$ is the difference between the predicted and observed distance moduli; $C$ is the corresponding covariance matrix that accounts for statistical and systematic uncertainties.\\

\noindent{\bf $BAO$ Data:} A combination of SDSS, 6dFGS, and Wiggle Z surveys at different redshifts have been utilised to gather the Baryon Acoustic Oscillations (BAO) data. The BAO data has been integrated with the followings:

\begin{equation}\label{eq.21}
d_{A}(z)=c \int_{0}^{z} \frac{d\tilde{z}}{H(\tilde{z})},
\end{equation}

\noindent where $d_{A}(z)$ is the comoving angular diameter distance, $c$ is the speed of light and $H(\tilde{z})$ represents the Hubble parameter. Also, the dilation scale can be obtained as, 
\begin{equation}\label{eq.22}
D_{v}(z)=\left[ \frac{d_{A}^2 (z) c z }{H(z)} \right]^{1/3},
\end{equation}

\noindent For the BAO data, the chi-square function ($\chi^{2}$) is employed, which can be expressed as
\begin{equation}\label{eq.23}
\chi^{2}_{BAO} = X^{T} C_{BAO}^{-1} X,
\end{equation}
where $C_{BAO}^{-1}$ represents the inverse covariance matrix and the determination of $X$ is on the specific survey being analyzed \cite{Giostri2012}. Using the above mentioned datasets, we have obtained the best-fit values of the parameters as summarized in TABLE \ref{table:I}.
\begin{table}[H]
\renewcommand\arraystretch{1.5}
\centering % used for centering table
\begin{tabular}{|c|c|c|c|} % centered columns (3 columns)
\hline\hline %inserts double horizontal lines
~~~Coeff.~~~&~~~~~~~ \textit{CC}  ~~~~~~~& ~~~~~~~\textit{SNIa}~~~~~~~ & ~~~~~$CC + SNIa + BAO$~~~~~\\ [0.5ex] % inserts table
%heading
% inserts single horizontal line
\hline\hline
$H_{0}$ & $69.01^{+0.11}_{-0.12}$ &  $68.99 \pm 0.09$ & $70.08_{-0.12}^{+0.10}$ \\
\hline
$\omega$ & $-1.03^{+0.11}_{-0.09}$ &  $-1.02^{+0.08}_{-0.09}$ & $-1.02 \pm 0.09$ \\
\hline
$\kappa$ & $-2.40 \pm 0.10$ &  $-3.22^{+0.13}_{-0.09}$ & $-3.19 \pm 0.10$ \\
\hline
$\lambda$ & $-0.01 \pm 0.08$ &  $-0.01^{+0.08}_{-0.11}$ & $-0.01 \pm 0.09 $ \\[0.5ex] % [1ex]
%[1ex] adds vertical space
\hline %inserts single line
$n$ & $1.38 \pm 0.09$ &  $1.29 \pm 0.09$ & $1.33 \pm 0.10$ \\[0.5ex]  
\hline %inserts single line
\end{tabular}
\caption{Marginalized constraints of the parameters using $Hubble$, $Pantheon$, and $BAO$ datasets.}
\label{table:I} 
\end{table}
%%%%%%%%%%%%%%
\begin{widetext}

\begin{figure}[H]
\centering
\includegraphics[width=12cm,height=12cm]{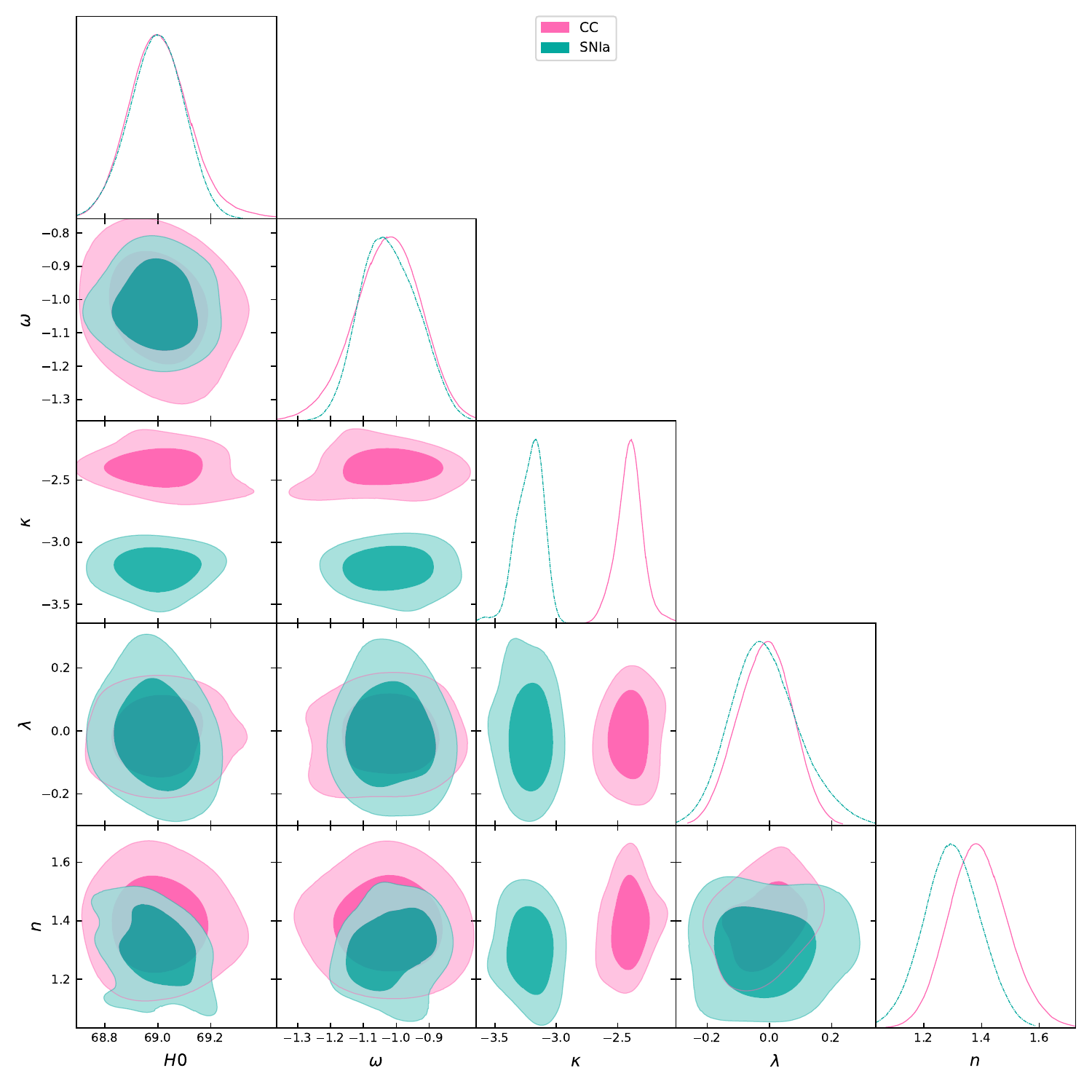}
\includegraphics[width=12cm,height=12cm]{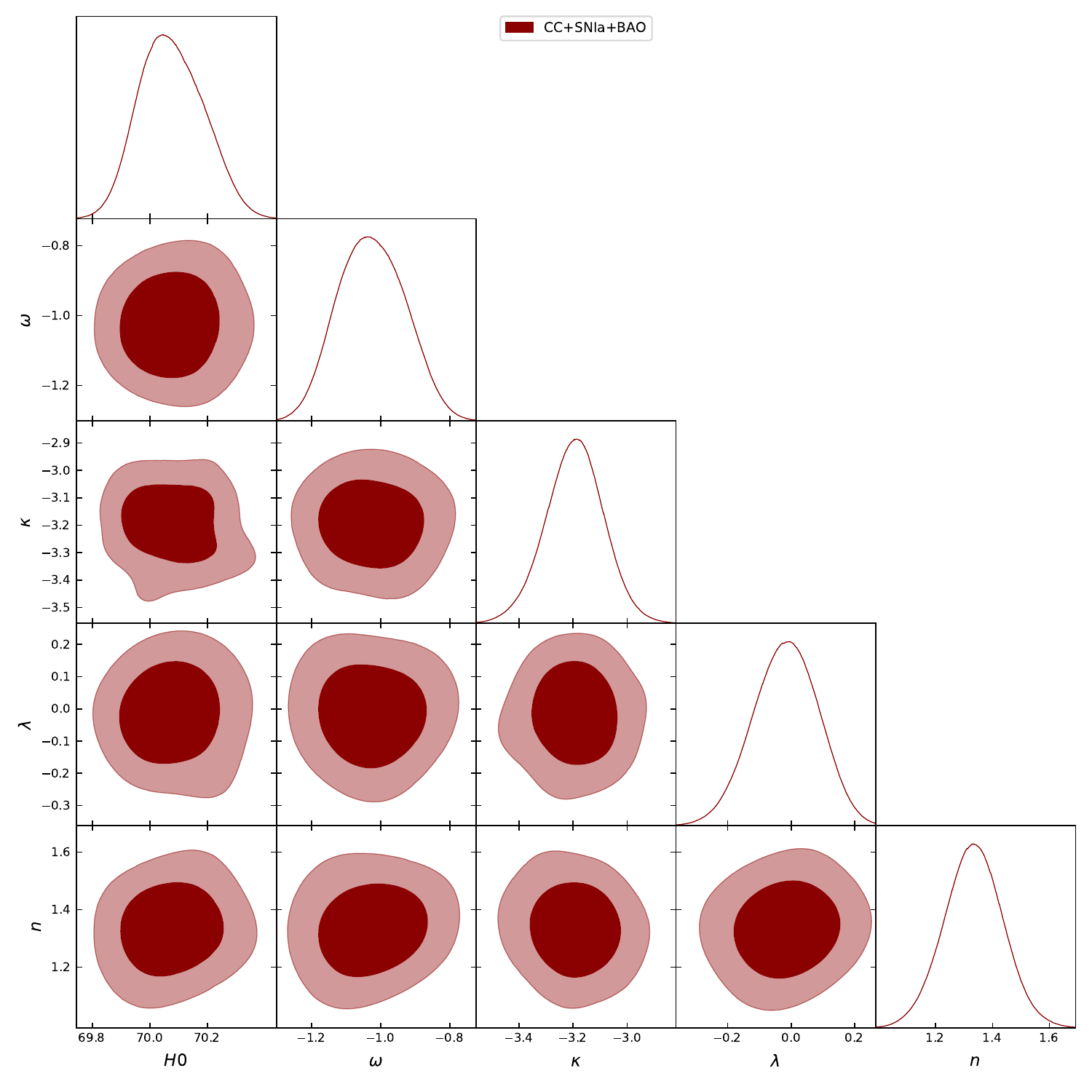}
\caption{The contour plots $CC$, $SNIa$ (\textbf{Upper panel}) and $CC+SNIa+BAO$ (\textbf{Lower panel}) with $1-\sigma$ and $2-\sigma$ errors for the parameters $H_{0}$, $\omega$, $\kappa$, $\lambda$ and $n$. Additionally, it contains the parameter values that better match the $32-$ points of $CC$ sample, $1048-$ light curves from $Pantheon$ dataset and $6-~BAO$ distance dataset.}
\label{fig:I}
\end{figure}

\begin{figure}[H]
\centering
\includegraphics[scale=0.5]{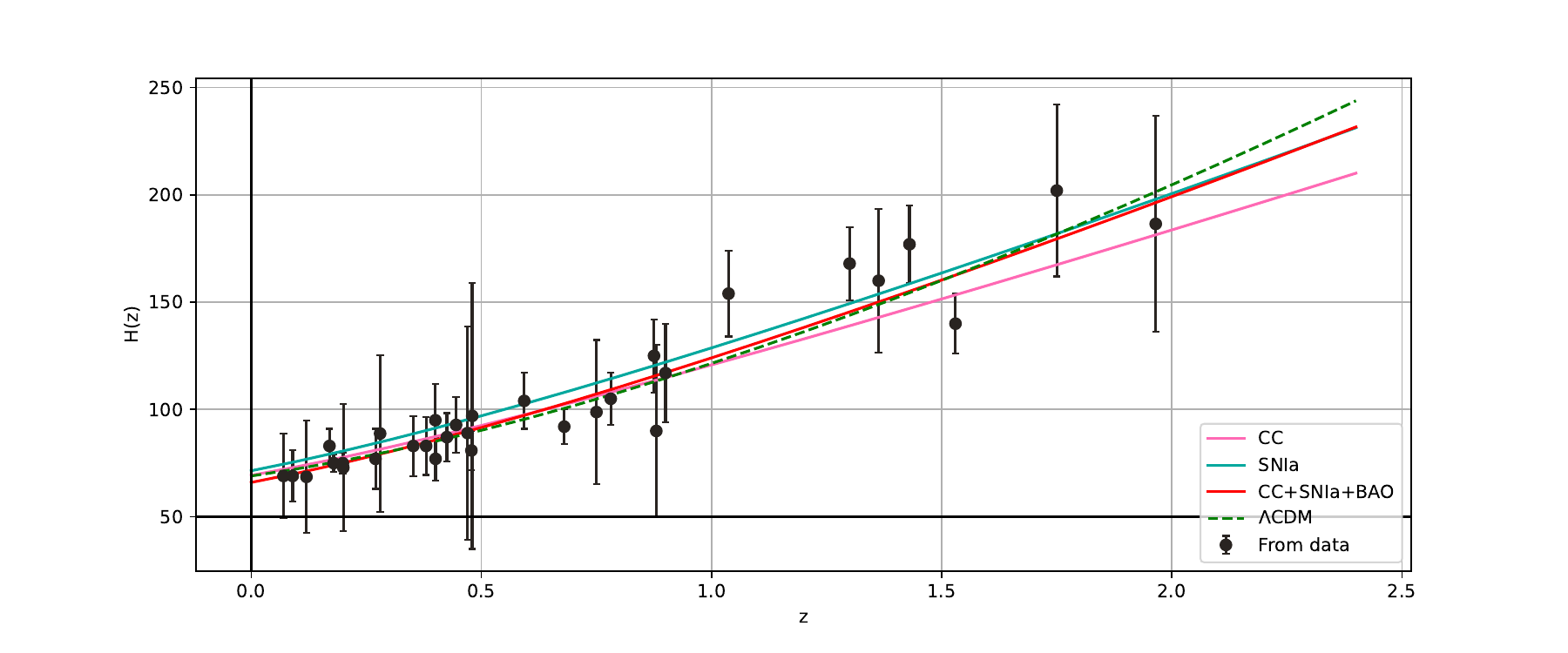}
\includegraphics[scale=0.5]{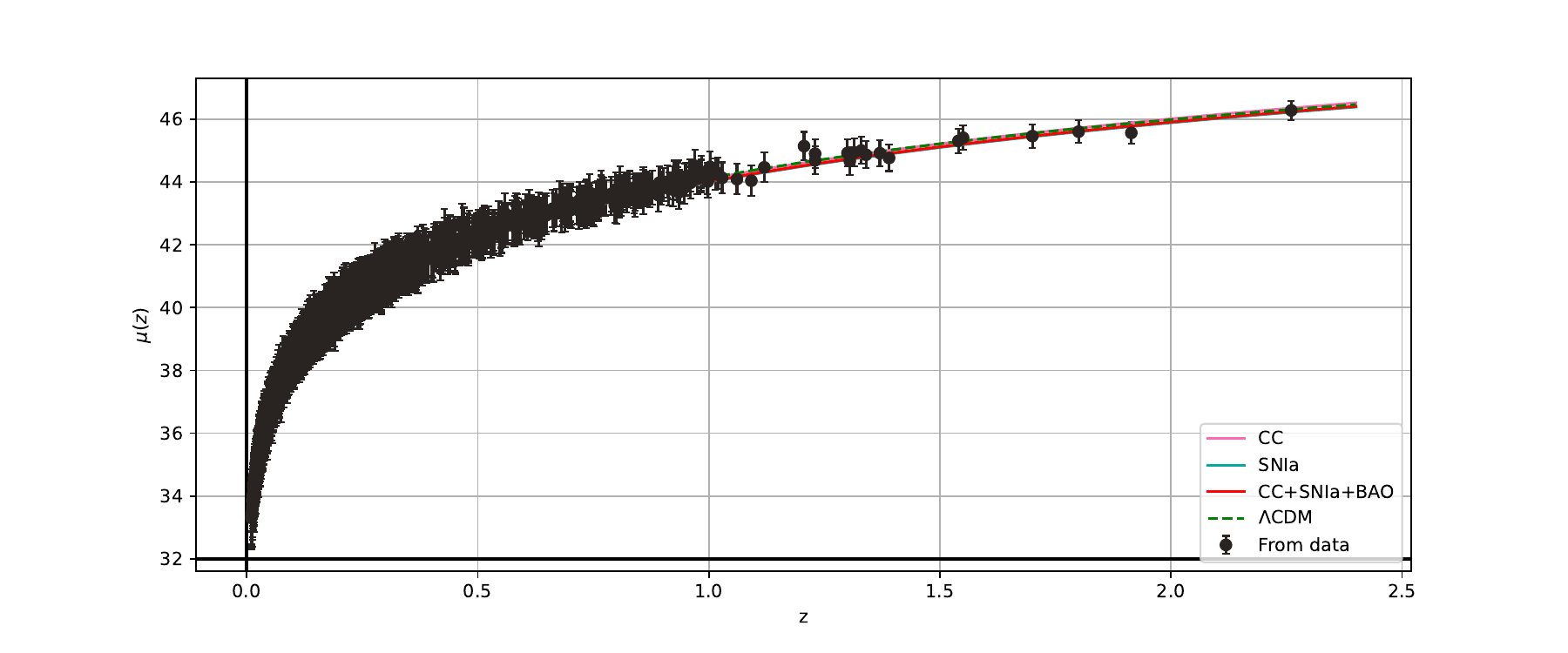}
\caption{Error bar for $CC$ sample (\textbf{Upper panel}), $SNIa$ dataset (\textbf{Lower panel}). The respective curve represent for the best-fit values obtained through different data sets.}
\label{fig:II}
\end{figure}
\end{widetext}
The contour plots for the two combinations of cosmological datasets, one with $CC$, $SNIa$ and the other one with $CC+SNIa+BAO$  are shown in FIG.- \ref{fig:I}. We have also presented the error bar plots using the $CC$ and $SNIa$ data and shown the curve pertaining our model and $\Lambda$CDM. It has been observed that the curve of the model traverse well withing the error bars [FIG.- \ref{fig:II}].  

\section{The Cosmographic Parameters and Energy Conditions}\label{Sec:IV}
From the Taylor series expansion of the scale factor, one can define the geometric parameters from the definition of cosmographic series as \cite{Capozziello2013},

\begin{eqnarray}\label{eq.24}
H &\equiv& \frac{1}{a}\frac{da}{dt},~~~ q \equiv -\frac{1}{aH^2}\frac{d^2a}{dt^2} \nonumber \\
j &\equiv& \frac{1}{aH^3}\frac{d^3a}{dt^3},~~~s \equiv\frac{1}{aH^4}\frac{d^4a}{dt^4}, 
\end{eqnarray} 

\begin{figure}[H]
\centering
\includegraphics[scale=0.5]{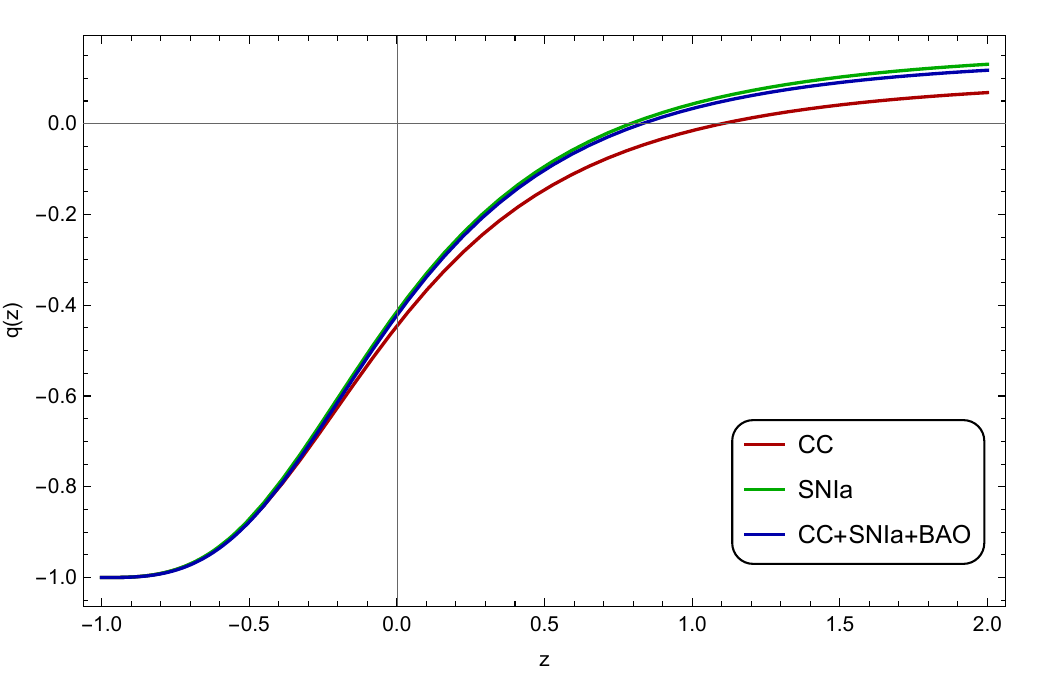}
\caption{Behaviour of deceleration parameter in redshift.}
\label{fig:q}
\end{figure}

The Hubble parameter $H$ plays a fundamental role in characterizing the evolution of Universe whereas the deceleration parameter provides information on the decelerating or accelerating behaviour of the Universe. The jerk and snap parameters provides further information on the behaviour of the model comparing to $\Lambda$CDM. Using Eq. \eqref{eq.17} and Eq. \eqref{eq.24}, we can express these parameters in terms of redshift as,
\begin{eqnarray}
q(z) &=& -1+\frac{(1+z)H_{z}(z)}{H(z)}, \nonumber\\
j(z) &=& q(z)\big[1+2q(z)\big] + (1+z)q_{z}(z), \nonumber\\
s(z) &=& \frac{j(z)-1}{3\left(q(z)-\frac{1}{2}\right)}, ~~~~~~~~~~~~~~\left( q\neq \frac{1}{2}\right), \label{eq.25}
\end{eqnarray}
where $H_{z}$ be the derivative of $H(z)$ with respect to $z$. The behavior of the Universe can be analysed by examining the present value ($q_0$) and and its range. To note, if $q_0 > 0$, the Universe experiences expansion but enters into the deceleration phase, which implies that the Universe is dominated by matter or a pressureless barotropic fluid. The present cosmological observations contradict the positive value of $q$ and it would have occurred at the early phase of evolution. If $q_0 < 0$, the Universe is expanding and accelerating, and if $q_0 = -1$, the energy content of the Universe is completely dominated by a de-Sitter fluid. FIG.- \ref{fig:q} shows the behaviour of deceleration parameter in different datasets. This illustrates the transition from deceleration to acceleration phase and the transition occurs at $z_t=1.1$, $z_t=0.79$, and $z_t=0.80$ respectively for the datasets of $CC$, $SNIa$, and $CC+SNIa+BAO$. The present value of $q$ has been listed in TABLE \ref{table:II}.

\begin{figure}[H]
\centering
\includegraphics[scale=0.7]{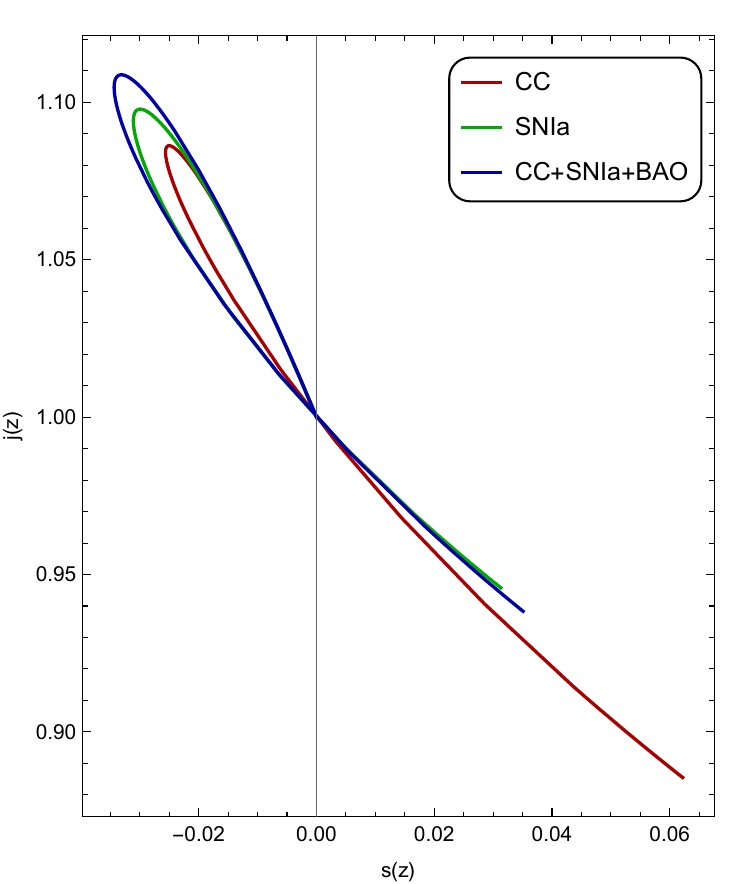}
\caption{Behaviour of state finder pair in redshift.}
\label{fig:js}
\end{figure}

From Eq. \eqref{eq.24}, the present value of the jerk parameter is, $j_0 = 2q_0^2 + q_0 + q_{z|_0}$. It must satisfy the condition, $2q_0^2 + q_0 > 0$ to keep $q_0$ in the range $[-1,-0.5]$. If $q_0 < -0.5$, the sign of $j_0$ is related to the variation of $q$. If $j_0<0$, the behavior of the Universe remains consistent from the current phase to the accelerated phase. If $j_0\rightarrow0$, the accelerating parameter smoothly converges to a specific value without any alteration in its behavior. If $j_0>0$, it indicates a distinct point during the evolution when the acceleration started, which is called the transition redshift ($z_{tr}$). The state finder pair $(j,s)$ characterizes the properties of dark energy in the model independent manner. Sahni et al. \cite{Sahni2003,Alam2003} have proposed this diagnostic with the classification: (i) $(j=1,~s=0)$ corresponds to the $\Lambda$CDM model, (ii) $(j<1,~s>0)$ indicates Quintessence, (iii) $(j>1,~s<0)$ represents Chaplygin Gas, and (iv) $(j=1,~s=1)$ signifies SCDM. So, the state finder pair helps to distinguish different dark energy models. The evolutionary behaviour of state finder pair for the constrained values of the free parameters has been shown in FIG.- \ref{fig:js} for the $CC$, $SNIa$ and $CC+SNIa+BAO$ datasets and the present value has been listed in TABLE \ref{table:II}. 

% The expanded expression of the cosmographic parameters [Eq.\eqref{eq.24}] are given in the \hyperref[Appendix]{Appendix}. 

\begin{table}[H]
\renewcommand\arraystretch{1.5}
\centering % used for centering table
\begin{tabular}{|c|c|c|c|} % centered columns (3 columns)
\hline\hline %inserts double horizontal lines
~~~Coeff.~~~&~~~~~~~ \textit{CC}  ~~~~~~~& ~~~~~~~\textit{SNIa}~~~~~~~ & ~~~~~$CC + SNIa + BAO$~~~~~\\ [0.5ex] % inserts table
%heading
% inserts single horizontal line
\hline\hline
$q_{0}$ & $-0.45$ &  $-0.42$ & $-0.42$ \\
\hline
$j_{0}$ & $1.03$ &  $1.06$ & $1.06$ \\
\hline
$s_{0}$ & $-0.01$ &  $-0.02$ & $-0.03$ \\
\hline 
\hline %inserts single line
\end{tabular}
\caption{Present value of cosmographic parameters.}
\label{table:II} 
\end{table}

\begin{figure}[H]
\centering
\includegraphics[scale=0.5]{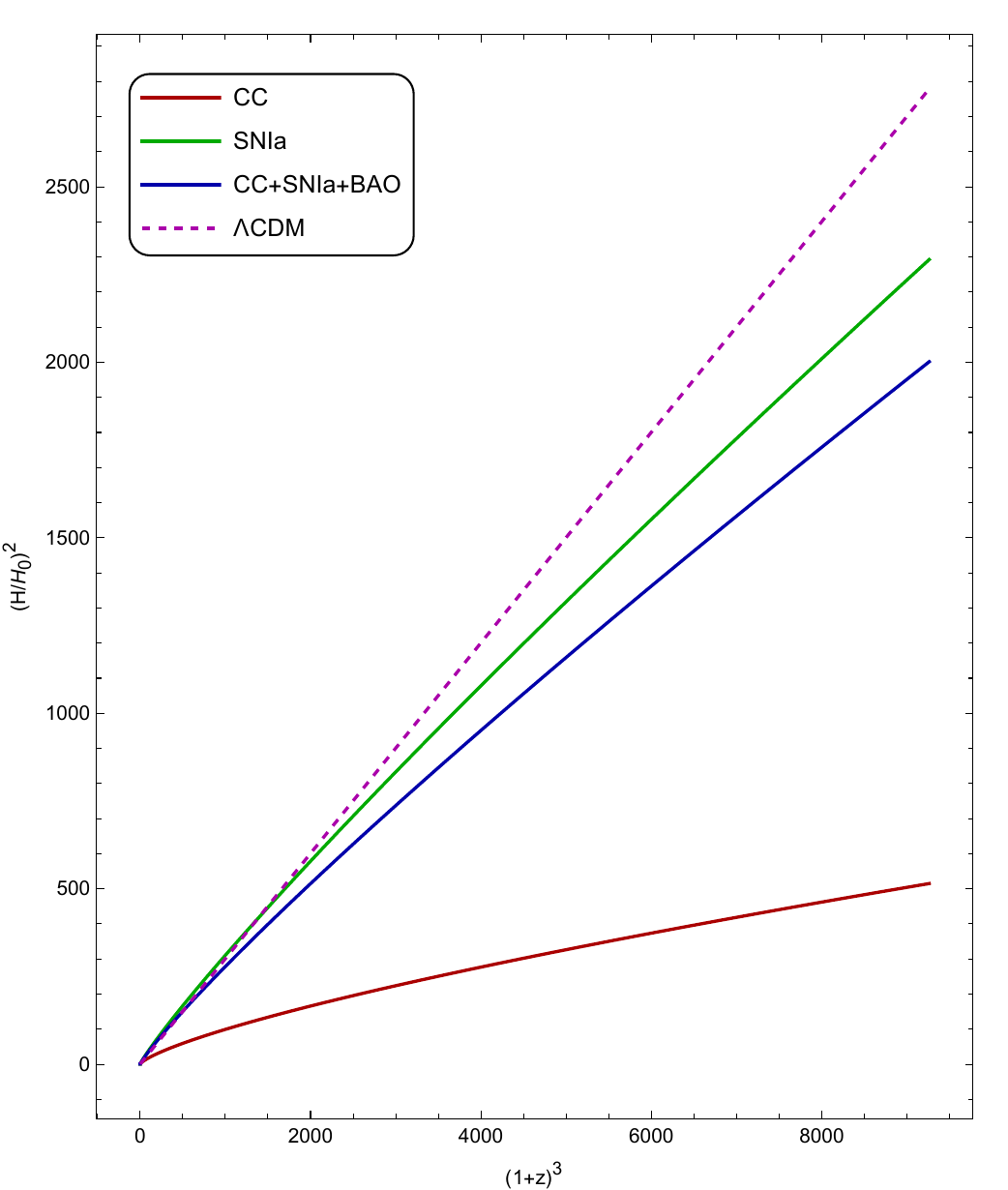}
\caption{Behaviour of $\big(\frac{H}{H_{0}}\big)^{2}$ versus $(1+z)^{3}$.}
\label{fig:omz}
\end{figure}

Another approach was suggested \cite{Sahni2008} to distinguish the $\Lambda$CDM model from other dark energy models without involving the EoS parameter. This approach is known as the  $Om(z)$ diagnostic. Several studies have provided the evidence for the sensitivity of the $Om(z)$ diagnostic to the EoS parameter \cite{Ding2015, Zheng2016, Qi2018}. The nature of the slope of $Om(z)$ differs among different dark energy models. A positive slope indicates the phantom phase ($\omega < -1$), while a negative slope indicates the quintessence region ($\omega > -1$). The $Om(z)$ diagnostic can be obtained as,

\begin{eqnarray}\label{eq.26}
Om(z) &=& \frac{\left[\frac{H(z)}{H_0}\right]^2 - 1}{(1+z)^{3}-1},\nonumber\\
&=& \frac{\left(-\frac{(1+z)^{52.7788 (\kappa +1) n (\omega +1)}+1}{\lambda  6^n (n+1) (2 n+1)-1}\right)^{1/n}-1}{(1+z)^3-1},
\end{eqnarray}
where $H_0$ is the present rate of the Hubble parameter. In FIG.- \ref{fig:omz}, the $\big(\frac{H}{H_{0}}\big)^{2}$ curve with respect to $(1+z)^{3}$ for $CC$, $SNIa$, $CC+SNIa+BAO$ datasets are below the $\Lambda$CDM curve, which confirms the phantom like behavior of model \cite{Sahni2008}.

Next we shall find the age of the Universe from our model. The expression for the age of the Universe is \cite{Vagnozzi2021},

\begin{equation}\label{eq.27}
t_u(z) = \int_{z}^{\infty} \frac{d\tilde{z}}{(1+\tilde{z})H(\tilde{z})}.
\end{equation}
Using Eq. \eqref{eq.17}, the expression can be obtained as,
\begin{eqnarray}\label{eq.28}
H_0(t-t_0) &=& \int_{0}^{z} \frac{d\tilde{z}}{(1+\tilde{z})E(\tilde{z})}, \nonumber\\
H_0t_0 &=& \lim_{z\to\infty} \int_{0}^{z} \frac{d\tilde{z}}{(1+\tilde{z})E(\tilde{z})}.
\end{eqnarray}

\begin{figure}[H]
\centering
\includegraphics[scale=0.5]{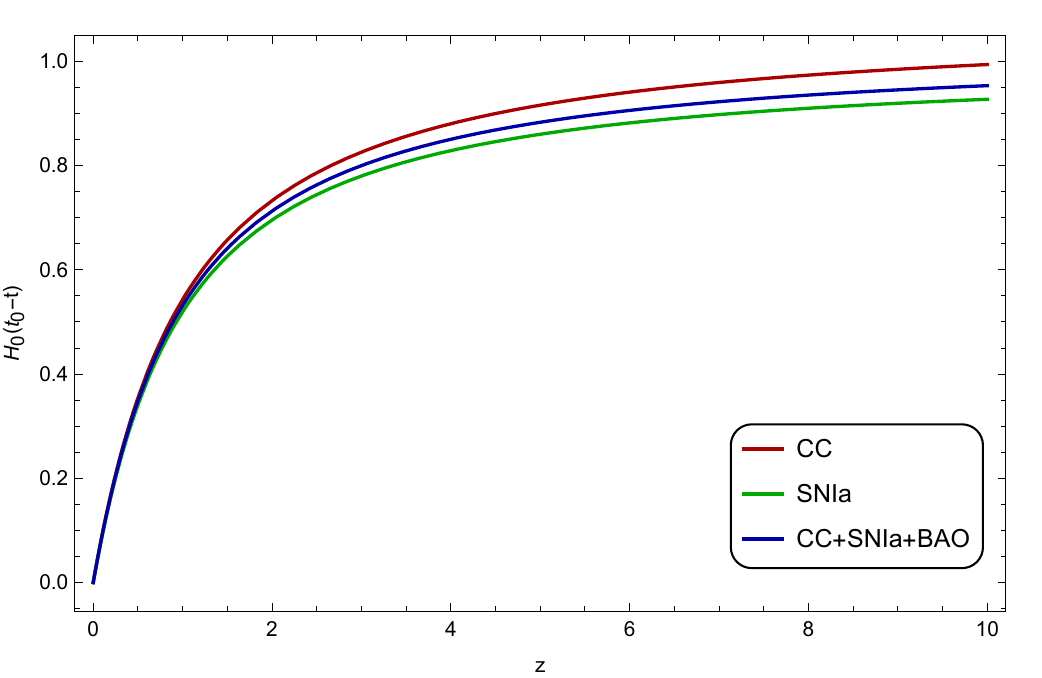}
\caption{Behaviour of $H_{0}(t_{0}-t)$ in redshift.}
\label{fig:aouz}
\end{figure}

 It is found that for infinitely large $z$, $ H_{0}(t_{0}-t)$ converges to $0.97$, $0.93$ and $0.95$ respectively for $CC$, $SNIa$, $CC+SNIa+BAO$ datasets [FIG.- \ref{fig:aouz}]. We can use this to calculate the current age of the Universe for respective dataset as $t_{0} = 14.05~~Gyrs$, $t_{0} = 13.48~~Gyrs$ and $t_{0} = 13.56~~Gyrs$. These values are in close agreement with the age determined from the Planck result, $t_0 = 13.78 \pm 0.020$ $Gyrs$. 

The energy conditions serve as the actual boundary conditions, which is an important aspect of the physical behaviour of the model. The energy density must be positive and because of the influence of dark energy, the limits of the boundary conditions may not exist. 
The energy conditions are weak energy condition: $\rho +p\geq0,~\rho\ge0$ (WEC); null energy condition: $\rho +p\geq0$ (NEC), dominant energy condition: $\rho -p\geq0$ (DEC), and strong energy conditions: $\rho+3p\geq0$ (SEC). We have given the expanded expressions in \hyperref[Appendix]{Appendix}.

From FIG.- \ref{fig:ec}, we can observe that the DEC satisfies through out the evolution and remains positive. Whereas the NEC violets, but tends to zero in late time. The violation of SEC due to the effect of DE, indicates that there is no physical reality to these boundary conditions. 
\begin{figure}
%\centering
\includegraphics[scale=0.5]{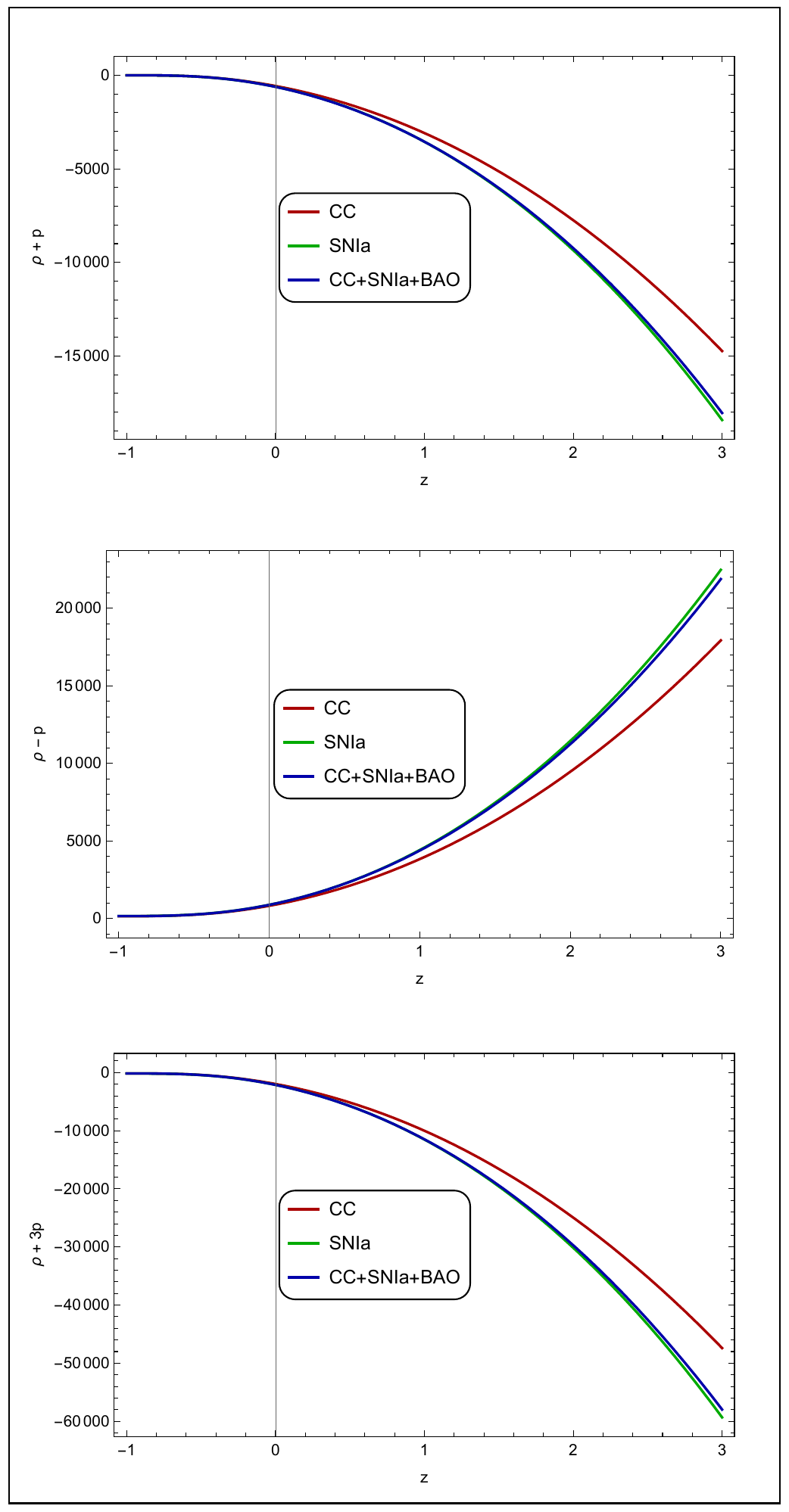}
\caption{Behaviour of $\rho+p$, $\rho-p$ and $\rho+3p$ in redshift.}
\label{fig:ec}
\end{figure}

\section{Conclusion}\label{Sec:V}
We have presented phantom cosmological model of the accelerating Universe in $f(Q,T)$ gravity, in which the non-metricity considered to be non-linear. Through some algebraic manipulation, we obtained an explicit expression for the Hubble parameter in terms of redshift. Using this expression, we parameterized $H(z)$ using the $Hubble$, $Pantheon$ and $BAO$ datasets. We focused on the redshift range $0.07 \leq z \leq 1.965$ and reconstructed $H(z)$ and the distance modulus using a $\chi^{2}$ minimization approach with $32$ data points. Also, we have used the Pantheon compilation data that includes $1048$ measurements of apparent magnitude for $SNIa$. In addition, we have incorporated $6-BAO$ points into our analysis. The MCMC analysis has been performed to obtain the best-fit values for the model parameters and EoS parameter (TABLE \ref{table:I}). From the error bar plots, we observe that the model curve and the $\Lambda$CDM curve pass through the range obtained from the datasets. 

Using the constrained values of the model parameters, We obtained the present value of the geometrical parameters such as the deceleration parameter, state finder pair and the $Om(z)$ diagnostic and analysed its behaviour. All these results provide the phantom behaviour of the model. The model exhibits a transition from a deceleration phase to an acceleration phase, occurring at a redshift of approximately for the datasets of $CC$, $SNIa$, and $CC+SNIa+BAO$ are $z_t=1.10$, $z_t=0.79$, and $z_t=0.80$ respectively. The age of the Universe obtained for different combination of the datasets and are similar that to of the recent cosmological observations results. The violation of SEC and non-violation of DEC have been obtained as expected in the context of modified theories of gravity. We can conclude that deviation from the $\Lambda$CDM model might indicates interactions between the dark energy and dark matter components. 

\section*{Acknowledgement}
RB acknowledges the financial support provided by University Grants Commission (UGC) through Junior Research Fellowship UGC-Ref. No.: 211610028858 to carry out the research work. BM acknowledges the support of IUCAA, Pune (India) through the visiting associateship program. 

\begin{widetext}

\begin{boxB}
 \section*{Appendix}\label{Appendix}

 \begin{eqnarray*}
     \rho &=& \frac{6 H^2 \left(\lambda  6^n H^{2 n}-1\right)-3 (5 \beta +48 \pi )H_{0}^2 \Omega_m (z+1)^3}{32 \pi  \left(\lambda  6^n (n+1) H^{2 n}-1\right)},\\
      p &=& -\frac{\lambda  2^{n+1} 3^n H^{2 n} \left(3 H^2-8 n (n+1) \dot H\right)-6 H^2+3 (7 \beta +48 \pi )H_{0}^2 \Omega_m (z+1)^3}{32 \pi  \left(\lambda  6^n (n+1) H^{2 n}-1\right)}.
  \end{eqnarray*}

 \begin{eqnarray*}
      \rho + p &=& \frac{\lambda  2^{n+2} 3^n n (n+1) H^{2 n} \dot H-9 (\beta +8 \pi )H_{0}^2 \Omega_m (z+1)^3}{8 \pi  \left(\lambda  6^n (n+1) H^{2 n}-1\right)},\\
     \rho - p &=& \frac{\lambda  2^{n+1} 3^n H^{2 n} \left(3 H^2-4 n (n+1) \dot H\right)-6 H^2+3 \beta H_{0}^2 \Omega_m (z+1)^3}{16 \pi  \left(\lambda  6^n (n+1) H^{2 n}-1\right)},\\
        \rho+3 p &=& -\frac{\lambda  6^{n+1} H^{2 n} \left(H^2-4 n (n+1) \dot H\right)-6 H^2+3 (13 \beta +96 \pi )H_{0}^2 \Omega_m (z+1)^3}{16 \pi  \left(\lambda  6^n (n+1) H^{2 n}-1\right)}.
    \end{eqnarray*}
\end{boxB}
 
\end{widetext}

\end{document}